\documentclass[runningheads]{llncs}
\usepackage[T1]{fontenc}
\usepackage{graphicx}
\usepackage[colorlinks=true,linkcolor=blue,urlcolor=blue,citecolor=blue]{hyperref}
\usepackage{comment}

\begin{document}
\title{Enhancing Network Security: A Hybrid Approach for Detection and Mitigation of Distributed Denial-of-Service Attacks Using Machine Learning}
\titlerunning{Hybrid Approach for DDoS Detection and Mitigation}
%
\authorrunning{N. J. Shohan, G. Tanbhir, F. Elahi, A. Ullah \& N. Sakib}

\author{Nizo Jaman Shohan\orcidID{0009-0008-7955-9785} \and
Gazi Tanbhir\orcidID{0009-0005-2584-6816} \and
Faria Elahi \and
Ahsan Ullah \and
Md. Nazmus Sakib}
%
%
\institute{Department of Computer Science and Engineering, World University of Bangladesh
\email{\{nizojamanshohan,gazitanbhir,fariae2019\}@gmail.com,
\{ahsan.ullah,nazmus.sakib\}@cse.wub.edu.bd}}
\maketitle              
\begin{abstract}
The distributed denial-of-service (DDoS) attack stands out as a highly formidable cyber threat, representing an advanced form of the denial-of-service (DoS) attack. A DDoS attack involves multiple computers working together to overwhelm a system, making it unavailable. On the other hand, a DoS attack is a one-on-one attempt to make a system or website inaccessible. Thus, it is crucial to construct an effective model for identifying various DDoS incidents. Although extensive research has focused on binary detection models for DDoS identification, they face challenges to adapt evolving threats, necessitating frequent updates. Whereas multiclass detection models offer a comprehensive defense against diverse DDoS attacks, ensuring adaptability in the ever-changing cyber threat landscape. In this paper, we propose a Hybrid Model to strengthen network security by combining the feature-extraction abilities of 1D Convolutional Neural Networks (CNNs) with the classification skills of Random Forest (RF) and Multi-layer Perceptron (MLP) classifiers. Using the CIC-DDoS2019 dataset, we perform multiclass classification of various DDoS attacks and conduct a comparative analysis of evaluation metrics for RF, MLP, and our proposed Hybrid Model. After analyzing the results, we draw meaningful conclusions and confirm the superiority of our Hybrid Model by performing thorough cross-validation. Additionally, we integrate our machine learning model with Snort, which provides a robust and adaptive solution for detecting and mitigating various DDoS attacks. 

\end{abstract}

\section{Introduction}
\begin{figure}[ht]
    \centering
    \includegraphics[width=0.8\linewidth]{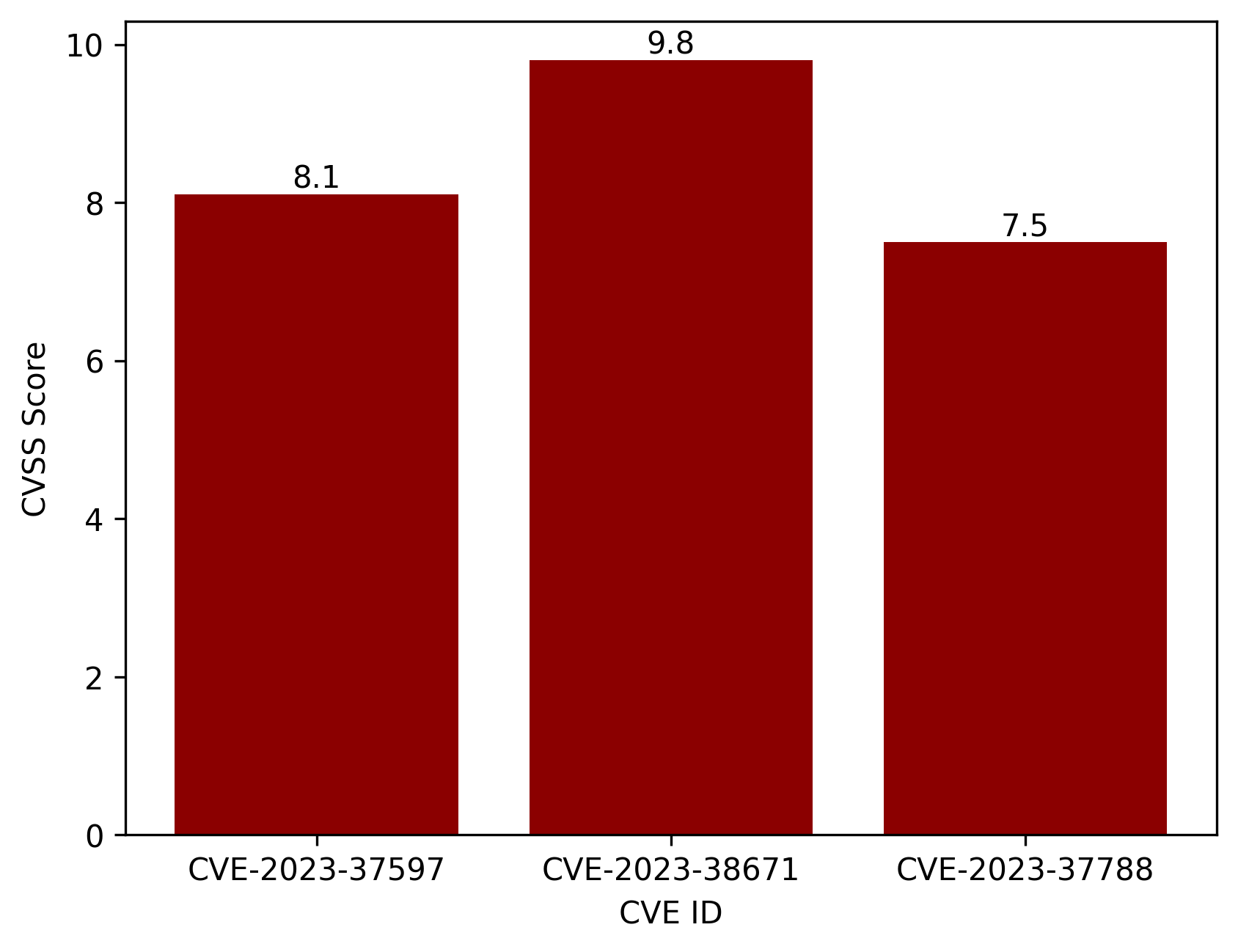}
    \caption{CVSS scores of various DDoS attacks in 2023\cite{CVE-2023-37597}\cite{CVE-2023-38671}\cite{CVE-2023-37788}.}
    \label{fig:cvss_scores}
\end{figure}

Distributed denial-of-service (DDoS) attacks pose a significant threat to the security of computer networks. A DDoS attack involves a deliberate effort to disturb the regular flow of traffic to a specific server, service, or network by inundating the target or its associated infrastructure with an excessive volume of Internet traffic\cite{def_DDoS}. DDoS attacks can wreak havoc in real life, disrupting essential online services and crippling businesses by rendering their websites and applications inaccessible. However, Fig.~\ref{fig:cvss_scores} vividly illustrates the distribution of CVSS Scores. CVSS, an acronym for the "Common Vulnerability Scoring System," represents a standardized framework within the field of cybersecurity. Its primary purpose is to evaluate and categorize the severity of vulnerabilities found within software, systems, and networks. It assigns a numerical score to vulnerabilities, ranging from 0.0 (least severe) to 10.0(most severe)\cite{CVSS_Paper}\cite{CVSS_NIST}.

Numerous researchers have historically employed traditional methodologies for detecting DDoS attacks, predominantly relying on signature matching. In this paradigm, the system compares incoming network traffic to a pre-existing database of known attack patterns or signatures\cite{szynkiewicz}. Despite its historical prevalence, this approach exhibits limitations when confronted with emerging and evolving attack types.

Despite the prevalent use of binary detection models by many researchers to identify DDoS incidents, these models may struggle to effectively adapt to evolving threat landscapes\cite{Kareem2022}\cite{Santhosh2023}. Due to their inability to recognize new attack patterns, frequent updates may be required to handle emerging DDoS techniques, rendering them less adaptable over time. In contrast, a multiclass detection model distinguishes attack types (e.g., UDP floods, SYN floods, HTTP floods), offering a comprehensive defense against diverse DDoS attacks. Its adaptability ensures a robust and sustainable defense in the ever-changing cyber threat landscape.

Based on these concerns, we proposed a Hybrid Model to bolster network security by synergizing the feature-extraction capabilities of 1D Convolutional Neural Networks (CNNs) with the classification prowess of Random Forest (RF) and Multi-layer Perceptron (MLP) classifiers. The study utilizes the CIC-DDoS2019 dataset for multiclass classification of various DDoS attacks\cite{Sharafaldin2019}. 

Finally, we integrate our machine learning model with Snort, a renowned Intrusion Detection and Prevention System (IDPS). This fusion of machine learning and Snort's Intrusion Prevention System (IPS) capabilities equip us with a robust and adaptive solution for detecting and mitigating DDoS attacks, enhancing the resilience of computer networks against evolving threats.

\section{Related Work}
Addressing the escalating threat posed by Distributed Denial of Service (DDoS) attacks, an advanced deep learning-based detection methodology is proposed. The approach utilizes Logistic Regression, K-Nearest Neighbor, and Random Forest algorithms with the NSL KDD dataset. However, limitations inherent in this dataset, notably its outdated nature and limited coverage of DDoS attacks, hinder a comprehensive analysis. Despite these constraints, the employed models exhibit marked enhancements in accuracy, particularly with the KNN and Random Forest models outperforming Logistic Regression. This underscores the imperative for future research to access more expansive and current datasets to bolster cybersecurity strategies, advocating for the development of real-time DDoS detection tools\cite{Pandey2023}.

In the landscape of cybersecurity, addressing Distributed Denial of Service (DDoS) attacks remains a critical concern prompting exploration of machine learning applications. However, existing research confronts challenges in achieving robust DDoS detection due to the attack's multifaceted nature. This study introduces a machine learning-based strategy employing XGBoost and Random Forest classifiers, utilizing the CICDDoS2019 dataset. While this approach demonstrates improved accuracy rates compared to previous methodologies, some drawbacks persist. Dataset selection profoundly influences model performance, and despite advancements, the modified XGBoost classifier showcases promising results but doesn't eliminate false positives entirely. Additionally, the training time for sophisticated classifiers like XGBoost remains considerable, limiting real-time applicability. Thus, while this study showcases progress in DDoS detection, challenges in dataset dependency and real-time implementation persist, warranting further refinement in future research endeavors\cite{Santhosh2023}.

Within the realm of IoT security, assessing the detection of DDoS attacks holds paramount importance, spotlighting the complexities and evasiveness inherent in these incursions. The use of Machine Learning (ML) stands as a key strategy in thwarting these threats, with researchers examining diverse ML algorithms through the lens of the CICDDoS2019 dataset. They advocate for enriching this dataset to effectively categorize a wider spectrum of attack types, underscoring the pivotal role of implementing novel algorithms to heighten the efficacy and efficiency of detection mechanisms\cite{Devi2023}.

The domain of network security and the detection and mitigation of DDoS attacks have undergone substantial advancements driven by pioneering research endeavors. Sanmorino and Yazid introduced the flow pattern-based DDoS attack detection by analysing network traffic characteristics, encompassing source and destination IP addresses, traffic types, and volume\cite{Sanmorino2013}.

Zekri et al. delved into the susceptibility of cloud computing environments to DDoS attacks and propose a DDoS detection system hinging on the C4.5 algorithm, underlining the potential of machine learning techniques in fortifying cloud resources\cite{zekri2017}.

Idhammad, Afdel, and Belouch introduced an online sequential semi-supervised machine learning approach for DDoS detection, adeptly amalgamating supervised and unsupervised techniques to elevate detection accuracy while curtailing false positives\cite{Idhammad2018}.

Mapanga et al. proposed a hybrid neural network design that combines MLP and CNN architectures in order to improve the detection rate of time-delayed assaults. This work focuses on detecting malicious nodes that deliberately or randomly drop packets destined for other target nodes. Furthermore, each packet drop attack is classified according to its attack type by watching and analyzing how each packet drop assault affects network properties. Furthermore, accomplishment in limiting false alarms during the detection of innovative assaults in the MANET environment IDS, spanning several packet dropping attack types such as selfish, sleep deprivation, and Blackhole attacks, is demonstrated\cite{Mapanga2017}.

The utilization of Snort and Zeek, integrated with Machine Learning (ML), within Software-Defined Networking (SDN) shows promise in distinguishing between benign and malicious traffic in simulated environments. However, a drawback lies within SDN itself, particularly in its susceptibility to DDoS attacks. Snort's focus primarily on volumetric DDoS attacks within SDN architecture might overlook non-volumetric variations, while also concentrating primarily on the control layer, potentially leaving other layers vulnerable. This highlights the necessity for future advancements in SDN security, encompassing a wider array of attack types and deeper analysis to fortify against diverse DDoS threats\cite{AbdulRaheem2023}. 

Furthermore, James et al. conducted an in-depth exploration of the effectiveness of 25 time-based features for the detection and categorization of 12 types of DDoS attacks. Their investigation encompassed both binary and multiclass classification. While achieving a notable 99\% accuracy in binary DDoS detection, the accuracy experienced a drop to approximately 70\% when classifying specific attack types\cite{Halladay2022}. In continuation of this foundational work, our research introduces a novel model that surpasses the 70\% accuracy achieved by existing models in multiclass detection.

\section{Methodology}
In this part the various steps of our Hybrid Model creation are given. Also, we described the detection and mitigation process of various DDoS attacks based on our hybrid approach.

\subsection{Preprocessing}
As mentioned, we have the CIC-DDoS2019 dataset, a standard benchmark dataset for our experimental results. We conducted a comprehensive check for missing values within the dataset using Pandas' isnull() function, which identifies missing values in dataset. After that, we dropped irrelevant columns, including one that served as the label for binary detection. Since the goal is to develop a multiclassification model, we retained only the column with multiclass labels.

\subsection{1D CNN Feature Extraction}
Due to its remarkable capacity for feature extraction, the utilization of 1D CNN is extensive in the analysis of time series data \cite{Kiranyaz2021}. In the initial stage, we partitioned the dataset into feature ($X$) and label ($y$) components to facilitate supervised machine learning. Subsequently, we divided the dataset into training and testing subsets, allocating 80\% for training and 20\% for testing, while ensuring reproducibility via a random seed of 42. Standardization of the features is performed using the StandardScaler() function from scikit-learn, thereby enhancing model convergence and reliability. 

\begin{figure}[ht]
    \centering
    \includegraphics[width=0.8\linewidth]{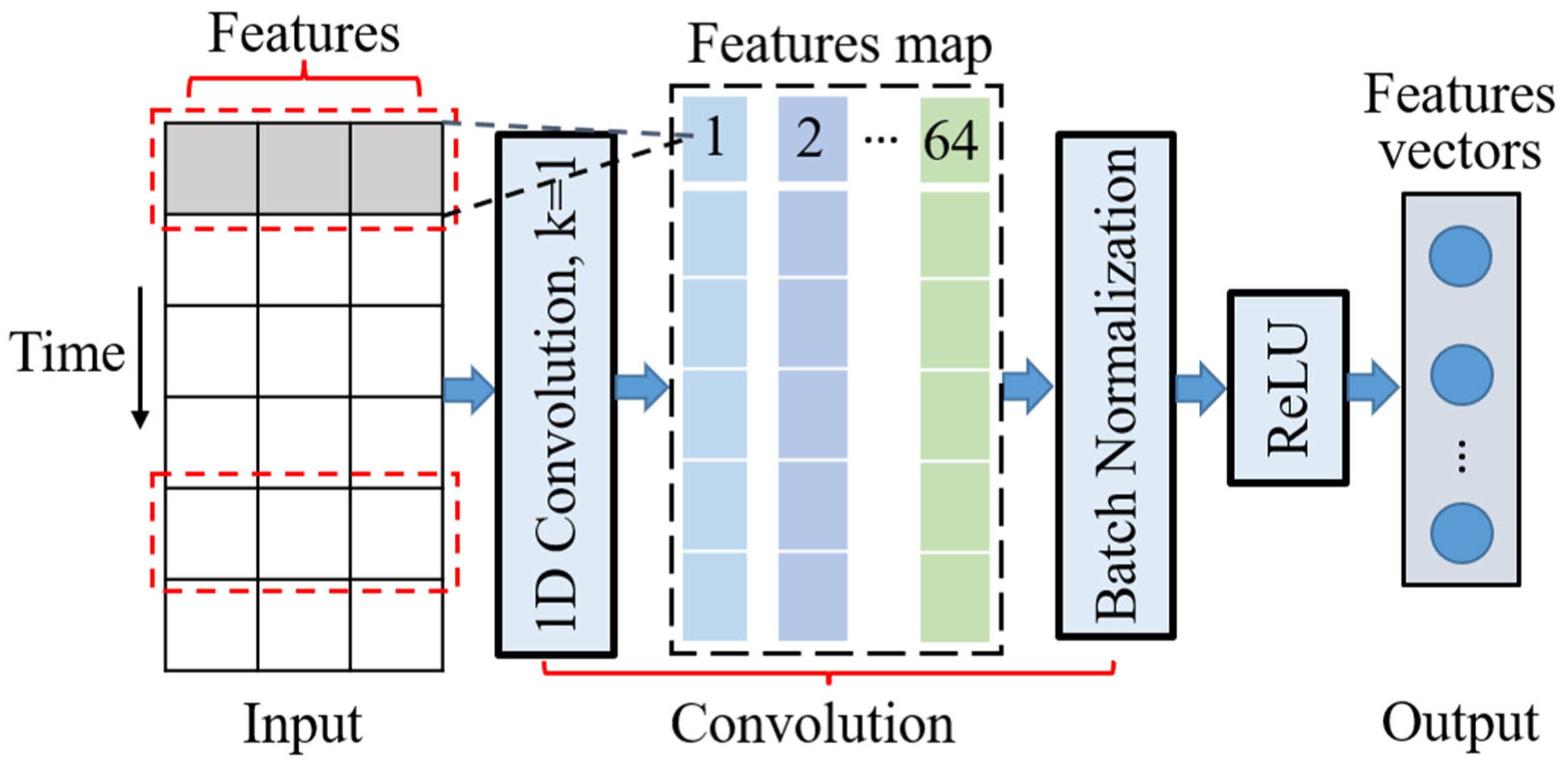}
    \caption{Feature extraction framework of 1D CNN\cite{Qin2023}.} 
    \label{fig:1dcnn}
\end{figure}

Furthermore, to align with the input requirements of our 1D CNN model, we reshaped the data into a 2D tensor using Python's reshape() method. This transformation ensured data compatibility for subsequent feature extraction. Our feature extraction methodology leveraged a 1D CNN model implemented with TensorFlow and Keras, comprising an input layer, a convolutional layer, and a global average pooling layer. The convolutional layer utilized filters and kernels to capture crucial data patterns, while the global average pooling operation summarized the extracted features, creating a representative feature set.

In 1D CNN, the convolution operation is a fundamental mathematical concept. It involves applying a convolutional kernel (a small filter) to the input signal or feature sequence. The kernel slides over the input and computes the dot product at each position, resulting in feature maps. Eq. (1) expresses the 1D convolution operation.

\begin{equation}
Y_i = \sum_{j=1}^N X_{i+j-1}W_j
\end{equation}

\noindent Note that in Eq. (1): \\
'$Y_i$' is the output feature map at position $i$.\\
'$X_{i+j-1}$' is the input sequence at position ${i+j-1}$.\\
'$W_j$' is the kernel's weight at position $j$.\\
'$N$' is the kernel size.\\

The Rectified Linear Unit (ReLU) activation function is commonly applied to the output of the convolution operation in convolutional neural networks to introduce non-linearity. Eq. (2) expresses the ReLU activation function.

\begin{equation}
ReLU(x) = \max(0, x)
\end{equation}

Then the feature extraction model is applied to the reshaped data, resulting in \texttt{X\_train\_features} and \texttt{X\_test\_features}.

\subsection{Random Forest Model Training}
Random Forest is an ensemble machine learning method that combines the predictions of multiple decision trees to enhance accuracy and reduce overfitting in classification and regression tasks. We leveraged the RandomForestClassifier from the scikit-learn library and trained our model using the X\_train\_features as input.

Gini impurity occurs during the training phase of the Random Forest model when we call \texttt{model\_rf.fit(X\_train\_features, y\_train)} and builds a forest of decision trees by making these feature and threshold decisions at each node on the training data. Eq. (3) expresses the Gini impurity.

\begin{equation}
Gini(p) = 1 - \sum_{i=1}^C p_i^2
\end{equation}

\noindent Note that in Eq. (3), 
$Gini(p)$ is the Gini impurity of the node.
$C$ is the number of classes.
$p_i$ represents the proportion of data points belonging to class $i$ in the node.

Having successfully trained the RF model, we proceeded to the testing phase. In this stage, the RF model made predictions on \texttt{X\_test\_Features} using \texttt{predict()} function and calculated the accuracy using \texttt{accuracy\_score()} function of the \texttt{sklearn.metrics} library.

Aggregating predictions by majority voting happens in the testing phase when we make predictions on the test data using the trained Random Forest classifier. Eq. (4) expresses the Majority voting.

\begin{equation}
\hat{y} = argmax_y \sum_{i=1}^n 1(\hat{y_i} = \hat{y})
\end{equation}

\noindent Note that in Eq. (4):\\ 
$\hat{y}$ is the final prediction.\\
$\hat{y_i}$ is the prediction of the $i$-th tree.\\
$n$ is the number of trees.\\
$1(\cdot)$ is the indicator function.

\subsection{MLP Model Training}
MLP is a foundational feedforward neural network architecture known for its capability to model complex non-linear relationships in data, making it a vital tool in various machine learning and pattern recognition tasks. During this training phase, we utilized the \texttt{MLPClassifier} from the scikit-learn library, and trained our model using X\_train\_features.

The MLP architecture consists of multiple layers, each with its set of neurons and weights. Eq. (5) expresses the output of a neuron in a hidden layer or the output layer.

\begin{equation}
y_j = \phi(v_j)
\end{equation}

\noindent Note that in Eq. (5):\\
$y_j$ represents the output of the $j$-th neuron in the layer.\\
$\phi(\cdot)$ represents the activation function.\\
$v_j$ represents the weighted sum of the input connections for the $j$-th neuron.\\

As an activation function, we used the ReLU activation function. Then, we proceeded to the testing phase. In this stage, the MLP model made predictions on \texttt{X\_test\_Features} as we did in the Random Forest model's testing phase.

The MLP learning process employs backpropagation, constituting a form of supervised learning. It involves adjusting the connection weights to minimize the error in the network's predictions. Eq. (6) expresses the error in an output node ($j$) for the $n$-th data point.

\begin{equation}
e_{j}{(n)} = d_{j}{(n)} - y_{j}{(n)}
\end{equation}

\noindent Note that in Eq. (6), 
$e_{j}{(n)}$ is the error at the $j$-th output node for the $n$-th data point.
$d_{j}{(n)}$ is the desired target value.
$y_{j}{(n)}$ is the actual output produced by the neuron.

The change in each weight ($w_{ij}$) can be computed using gradient descent. The weight update for a connection from neuron $i$ to neuron $j$ is given by:

\begin{equation}
\Delta w_{ij}{(n)} = -\eta \frac{\partial \varepsilon{(n)}}{\partial v_j{(n)}} y_i{(n)}
\end{equation}

\noindent Note that in Eq. (7), 
$\Delta w_{ij}{(n)}$ represents the change in weight from neuron $i$ to neuron $j$ for the $n$-th data point.
$\eta$ is the learning rate.
$\frac{\partial \varepsilon{(n)}}{\partial v_j{(n)}}$ is the gradient of the error with respect to the weighted sum of inputs for neuron $j$.
$y_i{(n)}$ represents the output of the previous neuron $i$.\\

These mathematical expressions underpin the learning process in the MLP, enabling it to adapt and improve its predictions over time.

\subsection{Proposed Hybrid Model}
\begin{figure}[ht]
    \centering
    \includegraphics[width=0.59\linewidth]{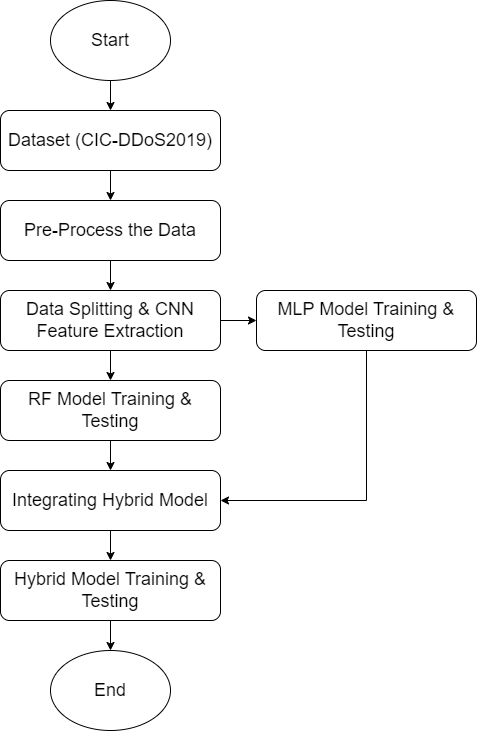}
    \caption{Integration of proposed Hybrid Model.} 
    \label{fig:hybrid_model}
\end{figure}
In Fig.~\ref{fig:hybrid_model}, we initiate our process by employing the CIC-DDoS2019 dataset. The journey begins with data preprocessing and then completes feature extraction using 1D Convolutional Neural Networks (CNNs). 

The workflow splits into two main paths: In the first path, we train the Random Forest model, evaluate its performance, and then proceed to the integration point for the Hybrid Model. Concurrently, the second path involves the training and performance assessment of the Multilayer Perceptron (MLP) model, culminating in a similar integration point for the Hybrid Model.

The next stage embodies the culmination of our research and development efforts. At this juncture, the Hybrid Model emerges as the result of our meticulous integration of the strengths of both the RandomForestClassifier and MLPClassifier. We utilized scikit-learn’s StackingClassifier to assemble both models and employ a meta-learner to make final predictions. And the final stage refers to the testing phase of our model.

\subsection{Integration of Hybrid Model with Snort}
To enhance network security, we introduced custom pre-processor that seamlessly integrate our machine learning-based Hybrid Model with Snort. We developed the custom pre-processor using custom C code to create the necessary pre-processing components. Our process began with defining functions within the custom pre-processor code responsible for packet processing, information extraction, and data transformation.

Subsequently, we integrated our custom pre-processor code with the Snort source code by modifying Snort's configuration files, effectively incorporating our custom pre-processor into Snort's pre-processing pipeline. After that, we recompiled Snort to include our custom pre-processor. This step involved running the appropriate compilation commands to rebuild the Snort executable, now equipped with our enhancements.

This process resulted in the development of a fortified security system, seamlessly blending the strengths of our Hybrid Model with Snort's robust mitigation capabilities.

\subsection{Detection and Mitigation Process}
\begin{figure} [ht]
    \centering
    \includegraphics[width=0.47\linewidth]{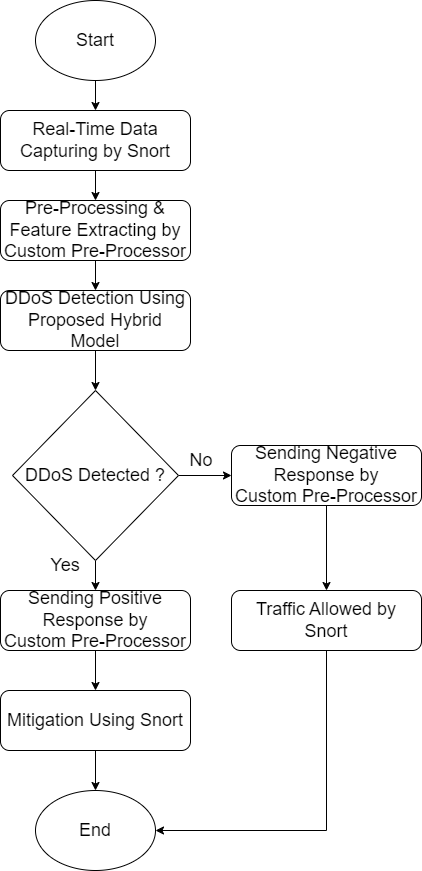}
    \caption{Proposed hybrid approach for detection and mitigation.} 
    \label{fig:detection_mitigation}
\end{figure}

In Fig.~\ref{fig:detection_mitigation}, real-time data is initially captured by Snort from an authorized source. The data goes through a custom pre-processing pipeline, which includes cleaning and formatting the captured traffic data, normalizing numerical features, and extracting relevant information for data transformation. The pre-processed data is then prepared for use with the Hybrid Model.

The Hybrid Model checks whether the traffic is malicious or not. If it identifies malicious characteristics, the pre-processor signals "Yes" to Snort, which proceeds with the mitigation process. If the data analysis does not indicate DDoS characteristics, Snort allows the traffic to pass.

\section{Experimental Results}
In this segment, we elucidate the proficiency and effectiveness of individual models to emphasize comparisons between the models on evaluation metrics. Furthermore, we confirm the superiority of our Hybrid Model by performing thorough cross-validation.

\subsection{Evaluation Metrics}
The following metrics are commonly used for evaluating classification models:

\begin{itemize}
    \item \textbf{Accuracy (ACC)} measures the proportion of correctly classified instances in the dataset. It is expressed as:
    \begin{equation}
        ACC = \frac{TP + TN}{TP + TN + FP + FN}
    \end{equation}
    
    \item \textbf{Precision (PRE)} quantifies the accuracy of positive predictions and is defined as:
    \begin{equation}
        PRE = \frac{TP}{TP + FP}
    \end{equation}
    
    \item \textbf{Recall (REC)}, also known as True Positive Rate (TPR) or Sensitivity, indicates the ability of the model to identify all relevant instances. It is given by:
    \begin{equation}
        REC = \frac{TP}{TP + FN}
    \end{equation}
    
    \item \textbf{F1 Score (F1)} is the harmonic mean of precision and recall and balances the trade-off between them:
    \begin{equation}
        F1 = 2 \cdot \frac{PRE \cdot REC}{PRE + REC}
    \end{equation}
\end{itemize}

\noindent Where: \\
$TP$ = Accurate Predictions of Positive Instances (True Positives) \\
$TN$ = Accurate Predictions of Negative Instances (True Negatives) \\
$FP$ = Inaccurate Predictions of Positive Instances (False Positives) \\
$FN$ = Inaccurate Predictions of Negative Instances (False Negatives) \\

\begin{figure}[ht]
  \centering
  \includegraphics[width=1\linewidth]{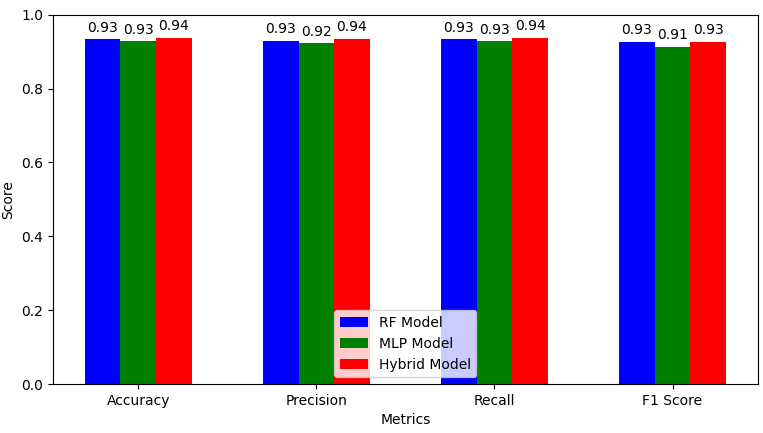}
      \caption{Performance comparison of individual models in evaluation metrics.}
  \label{fig:performance-comparison}
\end{figure}

These metrics provide a comprehensive view of the model's performance by considering different aspects of classification accuracy as Fig.~\ref{fig:performance-comparison}.

In our performance evaluation, the Hybrid Model, a combination of the Random Forest (RF) and Multi-Layer Perceptron (MLP) classifiers, outperformed both individual models. With an impressive overall average score of around 0.94, the Hybrid Model demonstrated superior accuracy and robustness compared to RF (0.93) and MLP (0.92). These results highlight the Hybrid Model's effectiveness, making it a compelling choice for multiclass predictive models across diverse applications.

\subsection{Cross-validation}
Cross-validation stands as a powerful tool for assessing a model’s generalization capabilities while identifying potential underfitting and overfitting concerns. Underfitting manifests when a model performs subpar on both training and validation data while overfitting is evidenced by the model excelling in training but faltering during validation in each fold. In contrast, a well-fitted model consistently demonstrates robust performance on both training and validation data.

\begin{figure}[ht]
  \centering
  \includegraphics[width=1\linewidth]{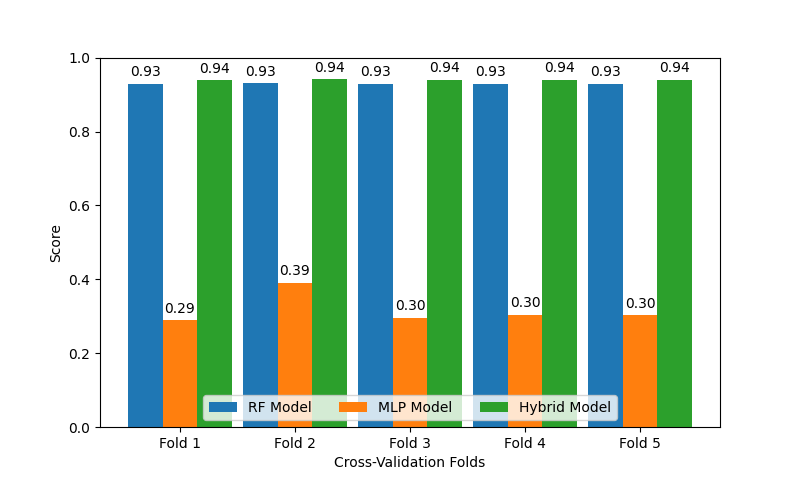}
      \caption{5-Fold Cross-validation for  individual models.}
  \label{fig:cv}
\end{figure}

In our case, the Hybrid Model consistently achieves accuracy scores of 0.94 in every fold during cross-validation shown in Fig.~\ref{fig:cv}, affirming its well-fitted nature and its ability to generalize effectively to new, unseen data. In comparison, while the Random Forest model also demonstrates strong performance with an accuracy of 0.93, there is a noticeable drop in accuracy for the Multi-Layer Perceptron model during cross-validation. This highlights the Hybrid Model's robustness and reliability in maintaining high accuracy across different folds, setting it apart in terms of performance and generalization capabilities.

\subsection{Discussion}
In the context of network security, the interplay between detection and mitigation is crucial. Improved detection forms the foundation for effective mitigation strategies. The integration of advanced machine learning models with Snort highlights the importance of enhancing the detection phase, which is critical for more effective mitigation.\\

\noindent The performance metrics of all models are summarized in Table 1.

\begin{table}[h]
    \centering
    \caption{Performance Metrics of All Models}
    \begin{tabular}{|l|c|c|c|c|c|}
    \hline
    & & & & & \\
    Models & Accuracy & Precision & Recall & F1 & Avg Cross-Val \\
    & & & & & \\
    \hline
    & & & & & \\
    RF Model & 0.93 & 0.93 & 0.93 & 0.93 & 0.93 \\
     & & & & & \\
     \hline
     & & & & & \\
    MLP Model & 0.93 & 0.92 & 0.93 & 0.91 & 0.31 \\
    & & & & & \\
     \hline
     & & & & & \\
    Hybrid Model & 0.94 & 0.94 & 0.94 & 0.93 & 0.94  \\
    & & & & & \\
    \hline
    \end{tabular}
\end{table}

\noindent These findings underscore the potential of our proposed Hybrid Model in enhancing network security, opening the door to more robust and accurate DDoS attack detection and mitigation. 

Although our research has produced promising results, a crucial aspect remains unexplored—namely, the real-world performance of our model. This uncharted territory stems from the complexities of conducting authentic DDoS attack simulations in controlled environments, where resource constraints, such as the requirement for multiple high-powered PCs, pose significant challenges.

\section{Conclusion}
This study introduces a robust Hybrid Model for DDoS attack detection, achieving an impressive 94\% accuracy in cross-validation and evaluation metrics — surpassing the 70\% accuracy of existing models in the case of multiclass detection. The model's capability for multiclass detection signifies a significant advancement in network security, emphasizes the shortcomings of binary models, and advocates for multiclass models in the ever-evolving threat landscape. It is essential to acknowledge the absence of real-world testing in the current study. Moving forward, we will focus on conducting rigorous real-world testing to validate the model's performance in authentic DDoS attack scenarios. This exploration is crucial for translating our promising findings into practical and impactful contributions to network security, ensuring the model's effectiveness in real-world applications.

%
%
%
%

\end{document}